\documentclass[preprint,12pt]{elsarticle}
\usepackage{placeins}
\usepackage{amssymb}
\usepackage{amsmath}
\usepackage{threeparttable}
\usepackage{booktabs}
\usepackage{xcolor}
\usepackage{graphicx}

\journal{Combustion and Flame}

\begin{document}

\begin{frontmatter}

\title{Quantitative 3D Morphology of Cellular H$_2$/O$_2$/N$_2$ Flames on a Porous-Plug Burner: Spatially Resolved Measurements of Temperature and OH Radical}

\author{Zeyu Yan, Xiangyu Nie, Shuoxun Zhang, Shengkai Wang*}

\affiliation{organization={SKLTCS, CAPT, School of Mechanics and Engineering Science, Peking University},
            addressline={5 Yiheyuan Road, Haidian District}, 
            city={Beijing},
            postcode={100871}, 
            country={China}}

\begin{abstract}
This study presents a systematic characterization of burner-stabilized lean hydrogen flame morphology across a wide range of equivalence ratios, dilution factors, and flow rates. Spatially resolved measurements of three-dimensional temperature and OH distributions were achieved. A comprehensive dataset of over 200 flame cases was obtained, enabling accurate determination of regime diagrams for different flame modes.  Linear stability analysis and direct numerical simulations were also performed and compared with the experimental results. The dominant wavenumbers of steady-state cellular flames were found to be consistently lower than the most unstable wavenumbers predicted by the linearized dispersion relation, indicating that nonlinear interactions between finite-amplitude perturbations of different length scales favored the growth of low-frequency components at long times. The cellular structures were found to be critically important in stabilizing the flame, especially at nominal equivalence ratios near the lean flammability limit. The mechanism of cellular flame stabilization was analyzed by complementary numerical simulations using a detailed reaction model. The combined effect of curvature-induced flame acceleration, local flow expansion/compression near the burner surface, and stratification of equivalence ratio caused by Soret diffusion created regions of reduced flow speed and enriched hydrogen concentration that helped anchor flames at nominal conditions where they would have blown off without the flame cells. The results of the present study are useful for understanding the fundamental flame dynamics of lean hydrogen mixtures and for improving the design of practical hydrogen combustors.
\end{abstract}


\begin{highlights}

\item Novelty and Significance\\
This work, to the authors’ knowledge, presents the first spatially resolved measurements of three-dimensional temperature and OH concentration fields in cellular flames of lean hydrogen mixtures. These measurements have quantified the ultimate statistics of burner-stabilized cellular flame morphology at steady state, which differ significantly from the results of classic linear stability analysis. The cellular structures were found to be critically important in stabilizing the flames, especially at nominal equivalence ratios near the lean flammability limit, where they would have blown off without the flame cells. Complementary numerical simulations have revealed the key mechanism of cellular flame stabilization. A systematic characterization of cellular flame instabilities across a wide range of equivalence ratios, dilution factors, and flow rates was also conducted, yielding a comprehensive dataset of flame modes, regime diagrams, and three-dimensional scalar distributions. The new findings and data obtained in the present study promise to advance both fundamental research on flame dynamics and practical applications of hydrogen combustion.

\item Author Contributions\\
Zeyu Yan: Data curation, Investigation, Writing - original draft. Xiangyu Nie: Data curation, Investigation, Writing - original draft. Shuoxun Zhang: Data curation,  Shengkai Wang: Conceptualization, Formal Analysis, Funding acquisition, Methodology, Supervision, Writing — original draft, Writing - review \& editing.
\end{highlights}

\begin{keyword}
Cellular Instability; Lean H$_2$ Flames; OH-PLIF; Thermometry
\end{keyword}

\end{frontmatter}

\section{Introduction}

The use of hydrogen as a carbon-free chemical energy carrier has attracted extensive research attention owing to its versatility, high specific energy, and non-toxic, clean nature \cite{pitsch2024transition}. Replacing conventional fossil fuels with hydrogen in current power generation, aviation, and industrial heating applications would enable substantial emission reductions and efficiency improvements, as well as increased operational flexibility \cite{verhelst2009hydrogen}. However, because hydrogen possesses thermodynamic, kinetic, and transport properties that are markedly different from those of conventional fuels, its combustion is often subject to intrinsic thermodiffusive instabilities, particularly under lean conditions commonly adopted in practical combustors \cite{matalon2007intrinsic, sanchez2014recent}. Under fuel-lean conditions, where the effective Lewis number is less than unity, the difference between local mass and heat transport rates can cause spontaneous wrinkling of the flame front and formation of cellular structures containing alternating enhancement and quenching patterns. These structures are important in the dynamic extinction and re-ignition processes of hydrogen flames \cite{robert2012thermal}. Moreover, they modify the overall flame propagation dynamics and the effective burning rate, posing challenges to the integration of hydrogen fuel into existing combustors optimized for hydrocarbon fuels. A comprehensive understanding of the cellular instability of hydrogen flames is therefore critically needed for the co-optimization of modern combustion devices and hydrogen-containing fuel streams.

Phenomenological observations of cellular flame instabilities have been well documented in the literature. For example, Chen and co-workers performed cinematographic recordings of cellular, non-premixed flame structures for a variety of fuel mixtures near their extinction limits \cite{chen1992diffusive}, and Robert and Monkewitz observed oscillatory cells in unstretched planar diffusion flames based on digital video measurements \cite{robert2012thermal}. More recently, Antar and co-workers conducted a series of chemiluminescence imaging studies of the reaction zones in cellular flames \cite{antar2023experimental, antar2024diffusive}. These studies have successfully identified the general range of flame conditions at which cellular instabilities occur; however, quantitative measurements of the flame morphology have been relatively scarce, with exceptions of planar laser-induced fluorescence (PLIF) imaging along selected cross sections of cellular flames. Representative examples include Jin and co-workers, who used OH-PLIF to image the cellular structure of premixed CH$_4$/H$_2$/air mixtures along the central vertical plane \cite{jin2015cellular, jin2015experimental}, and Jiang and co-workers, who conducted similar measurements in n-butane/air flames using OH-PLIF and CH$_2$O-PLIF \cite{jiang2020cellular, jiang2021experimental}. To date, however, spatially resolved three-dimensional measurements of key scalar distributions (for example, the temperature distribution) in cellular flames are still lacking. From a physics perspective, such measurements are particularly important because cellular flame instability is ultimately governed by the differential transport of scalars at microscopic scales.

On the computational side, there have been a multitude of high-fidelity numerical simulations of hydrogen cellular flames that focus on different aspects, including the linear dispersion relation of thermodiffusive instabilities \cite{altantzis2011detailed, frouzakis2015numerical, berger2019characteristic}, the nonlinear evolution of flame morphology \cite{kadowaki2005unstable, altantzis2012hydrodynamic, berger2022intrinsic2, berger2023flame}, the influence of Soret diffusion \cite{grcar2009soret}, and three-dimensional (3D) effects \cite{zirwes2024role, zirwes2024structure, wen2024thermodiffusively}. However, most of these studies are restricted to freely propagating flames, whereas realistic combustors invariably involve physical boundaries such as burner surfaces or confinement within an enclosure. Finite heat losses at these boundaries can substantially modify the flame instability characteristics, as previously observed in pulsating flames \cite{margolis1981effects, kurdyumov2008porous}, but high-resolution, comprehensive studies focused on cellular flames with boundary effects remain scarce.

Moreover, this boundary effect is often compounded by coupling with the flow velocity and the volumetric heat-release rate, which together modulate the effective temperature gradient at the thermal boundary by affecting the flame standoff distance and the post-flame temperature, respectively. Previous studies have focused mainly on premixed fuel-air flames, where volumetric heat-release rates are relatively low (e.g. \cite{jin2015experimental, jiang2020cellular, jiang2021experimental, nakatsuru2021shape}). Flames with enriched or pure oxygen as the oxidizer are expected to exhibit substantially different instability boundaries, but this remains to be explored further as well.

In light of these issues, the present study aims to systematically characterize burner-stabilized cellular flames of lean hydrogen mixtures across a wide range of equivalence ratios, dilution factors, and flow rates, based on spatially resolved measurements of three-dimensional temperature and OH distributions. Complementary numerical simulations are also performed under representative cellular flame conditions to analyze the effects of differential diffusion and heat loss on the morphology and stabilization mechanism of the cell structures. The remainder of this paper is organized as follows. Section 2 elaborates on the experimental and computational methods used in the current study. Section 3 discusses the results, including measured scalar distributions, instability regime diagrams, and dominant wavenumbers obtained from a collection of over 200 lean hydrogen flames, as well as representative numerical simulations regarding the details of cell structure. Section 4 concludes this paper with a summary of the main findings and an outlook.

\section{Methods}
\subsection{Experimental Measurement of 3D Temperature and OH Distributions}
\begin{figure}[ht!]
\centering
\includegraphics[width=\columnwidth,trim = 0 0 -2 0,clip]{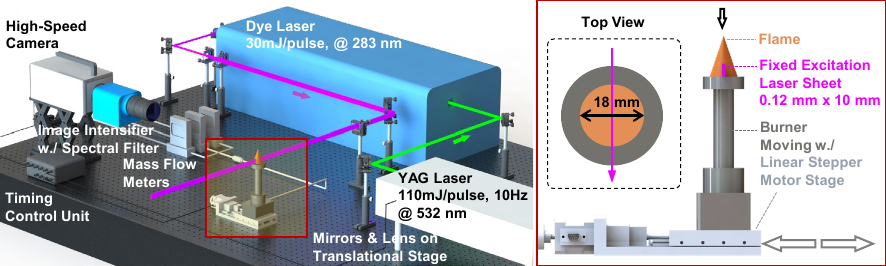}
\caption{Schematic of the current experimental setup.}
\label{fig1}
\end{figure}

A schematic view of the current experimental configuration is shown in Fig. \ref{fig1}. The flame experiments were conducted on a custom-built McKenna-type axial symmetric porous-plug burner. The porous-plug burner generated a relatively uniform incoming flow, thereby mitigating the effect of strain on the onset of flame instability. Heat loss across the burner surface also helped stabilize the cellular structure, enabling detailed measurement over extended times. Similar porous-plug burners have been used in previous experimental studies of the premixed flame cellular instability for other fuels, such as CH$_4$ \cite{konnov2005measurement} and n-C$_4$H$_{10}$ \cite{jiang2020cellular}. 

As illustrated in Fig. 2, the burner used in the current study featured a circular sintered bronze plug of 18 mm diameter that was water-cooled to maintain a stable temperature. The temperature of the porous plug was continuously monitored using a K-type thermocouple (Omega TJ36-CAXL-020-12) embedded approximately 2 mm beneath the burner surface. High-purity fuel (99.99\%-grade H$_2$), O$_2$ (99.999\%-grade), and N$_2$ (99.999\%-grade) were supplied to the bottom of the burner, where they were thoroughly mixed by an in-line static mixer before passing through the porous plug. The flow rates of H$_2$, $\rm O_2$, and $\rm N_2$ were precisely controlled by three Alicat MC-series mass flow controllers, with typical uncertainties of 0.1\%, 0.2\%, and 0.5\%, respectively.

\begin{figure}[ht!]
\centering
\includegraphics[width=0.5\columnwidth]{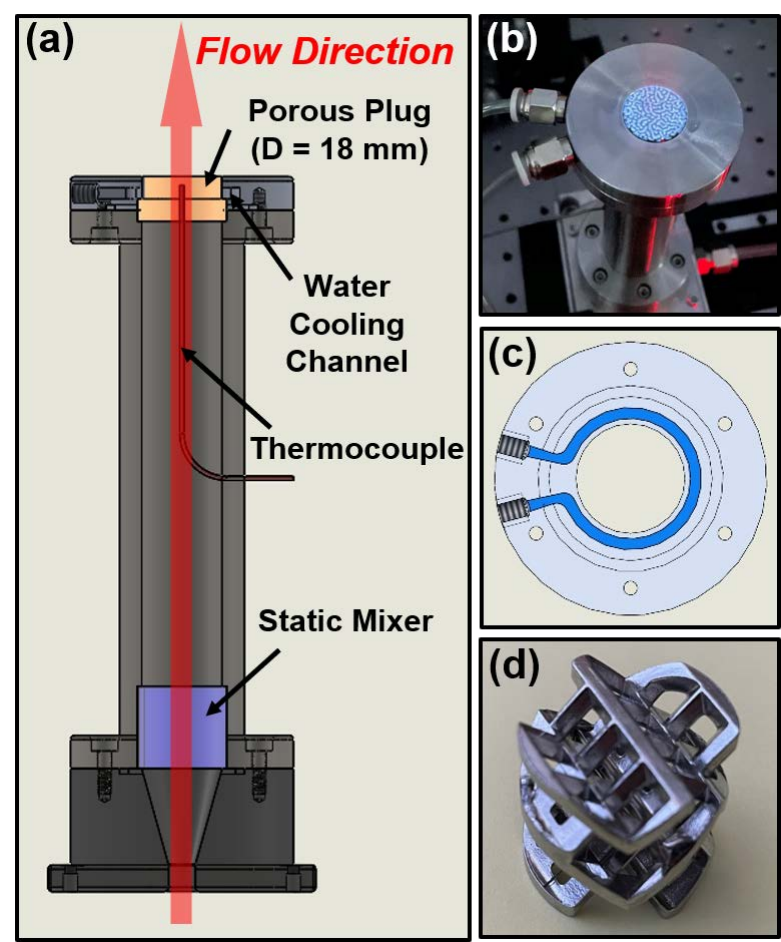}
\caption{Configuration of the porous-plug burner used in the current experiment. (a) Cross-sectional view of the burner; (b) example of a cellular flame; (c) details of the water cooling channel; (d) photograph of the static mixer.}
\label{fig2}
\end{figure}

Spatially resolved measurements of the cellular flame structure were performed using OH-PLIF. The A-X (1,0) P(1.5)+Q(1.5)+R(2.5) transition cluster near a vacuum wavelength of 282.997 nm and the R(9.5) transition of OH at 282.952 nm were excited by a ns-pulsed tunable dye laser (LIOP-TEC, Model LiopStar-N) at a repetition rate of 10 Hz. The use of two wavelengths enabled determination of gas temperature from their intensity ratio, as demonstrated in a previous study by the authors \cite{wang2019quantitative}. The dye laser probing these transitions was optically pumped by a frequency-doubled pulsed Nd:YAG laser (InnoLas, Model SpitLight2000-10) at 532 nm to generate coherent radiation at 566 nm, which was then passed through a frequency-doubling module to produce 30-mJ UV pulses near 283 nm. The laser pulse energy was measured by a Newport power meter (Model 843-R) with an uncertainty of less than 5\%. The excitation laser beam was expanded by three cylindrical fused silica lenses (with focal lengths of -25 mm, 500 mm and 500 mm, respectively) into a thin sheet of 0.12 mm thickness (2-$\sigma$ value), illuminating a region of approximately 10 mm height. The OH-PILF signal was bandpass-filtered at 310 ± 10 nm before being collected by an EyeiTS image intensifier and recorded by a Phantom v611 high-speed digital camera with a pixel resolution of 256 $\times$ 512. The physical size of an individual pixel was determined to be 88 $\mu$m $\times$ 88 $\mu$m using a geometry calibration target. The effective spatial resolution, usually limited by the image intensifier,  was determined to be 0.28 mm, on the order of 3 pixels (see Supplementary Material). Further details on the laser diagnostic setup have been documented in the authors' previous work \cite{wang2025self} and are omitted here for brevity.

To fully resolve the 3-D cellular structure, a linear sweep of the excitation laser sheet across the hydrogen flame was performed. Specifically, the porous burner was placed on an electrically driven linear translation stage that moved continuously at a speed of 0.2 mm/s during each measurement, whereas the absolute position of the laser sheet remained unchanged. The motion of the porous burner was synchronized with the excitation laser, image intensifier, and high-speed camera using a central timing unit. The exposure time of the intensified camera was set to be 10 ns -- long enough to cover the pulse width of the laser and short enough to minimize the influence of background emission. The intensifier gain was adjusted to yield a maximum pixel value of about 75\% of the camera's full range in order to avoid saturation effects. 

\begin{figure}[ht!]
\centering
\includegraphics[width=0.5\columnwidth]{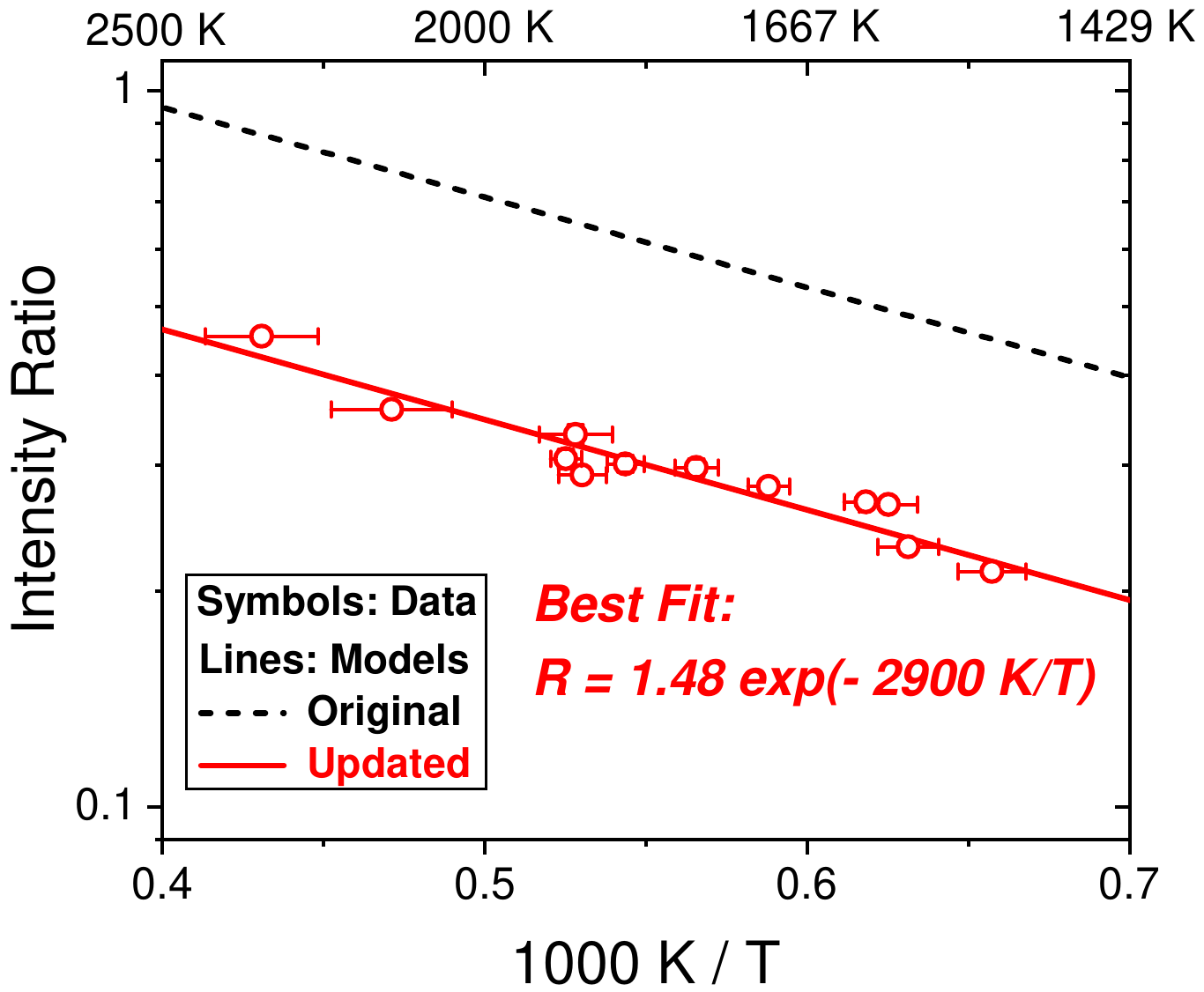}
\caption{Temperature dependence of the intensity ratio between PLIF signals excited at 282.952 nm and 282.997 nm. Symbols: current measurement results; dashed line: calculation using a recent spectroscopy model \cite{yousefi2018new}; red solid line: updated version of the model that maintains the same slope as the original but with the pre-exponential factor adjusted to best fit the experimental data.} 
\label{fig3}
\end{figure}

The measured PLIF signal was normalized by the excitation laser intensity distribution obtained from an acetone calibration experiment. The temperature distribution of a cellular flame was then inferred from the ratio of normalized PLIF signals at two excitation wavelengths, based on an empirical relation shown in Fig. \ref{fig3}. This relation was determined from a series of thermometry calibration experiments performed at stoichiometric conditions, where thermodiffusive instabilities were absent and the flame front was stable and relatively uniform. In the calibration experiments, the gas temperature distributions were measured independently using a spatially-resolved laser absorption method developed in a previous study of the authors \cite{zhang2025precision}. The average values of the gas temperature and the normalized PLIF signal in a 4 mm × 4 mm core region behind the flame front were used to calculate an empirical fit of the PLIF intensity ratio ($R$) as a function of temperature ($T$): $R = 1.48 \rm{exp}(-2900 K/\textit{T})$. Note that the exponential term in this expression corresponds to the lower-state energy difference between the two excitation transition clusters, which is relatively well documented in the literature. In the current study, its value was calculated based on a recent OH line list reported by Yousefi and co-workers \cite{yousefi2018new}. The pre-exponential factor, on the other hand, carried more uncertainty and was determined directly from the current numerical fit. Further details of the calibration experiments can be found in the Supplementary Material. 

Once the gas temperature distribution was obtained, the OH concentration was determined from the PLIF signal excited at 282.997 nm by properly accounting for fluorescence quenching using a detailed spectroscopy model named StaR-LIF \cite{yan2025star}. Specifically, the fluorescence signal $S$ was modeled as the product of the OH mole fraction $X_{OH}$, laser intensity $I$, gas pressure $P$, effective volume of measurement $V$, solid angle of collection $\Omega$, absorption coefficient $k_{OH}$, fluorescence quantum yield $FQY$, and a transfer function $\eta$ representing the overall efficiency of fluorescence collection and imaging optics, as illustrated in Eqn. (1).
\begin{equation}
\begin{aligned}
S = X_{OH} \cdot I \cdot P \cdot V \cdot \frac{\Omega}{4\pi} \cdot k_{OH} ( T, P, \{X_i\})\cdot FQY(T,P,\{X_i\}) \cdot \eta\\
\end{aligned}
\end{equation}

In the current study, $P, V, \Omega, \eta$ remained the same throughout all experiments, and the effect of $I$ was already accounted for via intensity normalization. This leads to a simplified equation as follows:

\begin{equation}
\begin{aligned}
S = C \cdot \psi(T,\{X_i\}) \cdot X_{OH}
\end{aligned}
\end{equation}

In Eqn. (2), the effects of all constant parameters are lumped into a single proportional factor $C$, and the influence of gas temperature and composition (calculated with Cantera \cite{cantera}) is represented by a nondimensional function $\psi$. In the current study, the value of $C$ was determined by comparing the measured and simulated PLIF signal in a homogeneous region of a representative stable flame (at the conditions of $\dot{m}_{\rm H_2}$ = 1.00 SLPM, $\dot{m}_{\rm N_2}$  = 1.00 SLPM, and $\dot{m}_{\rm O_2}$ = 2.59 SLPM), whereas $\psi$ was calculated on a case-by-case basis using the StaR-LIF model \cite{yan2025star}.

\subsection{Theoretical Analysis of the Linearized Instability Growth Rates}
In the present work, the wavenumber dependence of the linearized instability growth rate, i.e. the dispersion relation, was analyzed under both adiabatic and non-adiabatic conditions, based on theoretical models developed previously by Sivashinsky \cite{Sivashinsky1977nonlinear} and by Jin and co-workers \cite{jin2015experimental}, respectively. A unified form of the models, adapted from \cite{jin2015experimental}, is as follows:

\begin{equation}
\begin{aligned}
\omega(k) & = \frac{\sqrt{\sigma^3+\sigma^2-\sigma-(\sigma^2-1)(g/S_L^2k)}-\sigma}{\sigma(\sigma+1)} S_L k \\
& +\frac{H+Ze(1-Le)/2-1}{1-H} \sigma D_T k^2-\frac{4-H}{1-H}\frac{D_T^3}{S_L^2} k^4
\end{aligned}
\end{equation}

In this equation, $\omega(k)$ denotes the growth rate of a sinusoidal instability mode of wavenumber $k$, $\sigma = \rho_u/\rho_b$ is the expansion ratio, $S_L$ is the laminar flame speed, $Ze$ is the Zeldovich number representing the non-dimensional activation energy, $Le$ is the effective Lewis number of the reactive mixture, $D_T$ is the thermal diffusivity, and $H = E_A/R\cdot(1/T_b-1/T_{b,ad})$ is a non-dimensional factor representing the heat loss effect. In the absence of heat loss ($H=0$), Eqn. (3) is equivalent to the dispersion relation proposed by Sivashinsky \cite{Sivashinsky1977nonlinear}. In the current study, the value of $H$ was determined directly from the temperature measurement data. For cellular flames, $T_b$ was calculated from the average value in a relatively uniform region sufficiently far from the flame front, whereas $T_{b,ad}$ was obtained from Cantera simulation, and the activation energy was determined from $E_A/R = -2\rm{d ln}(\rho_u S_L)/d(1/T_b)$ by perturbing the unburned gas temperature $T_u$. 

\begin{figure}[ht!]
\centering
\includegraphics[width=1\columnwidth,trim = 0 0 -2 0,clip]{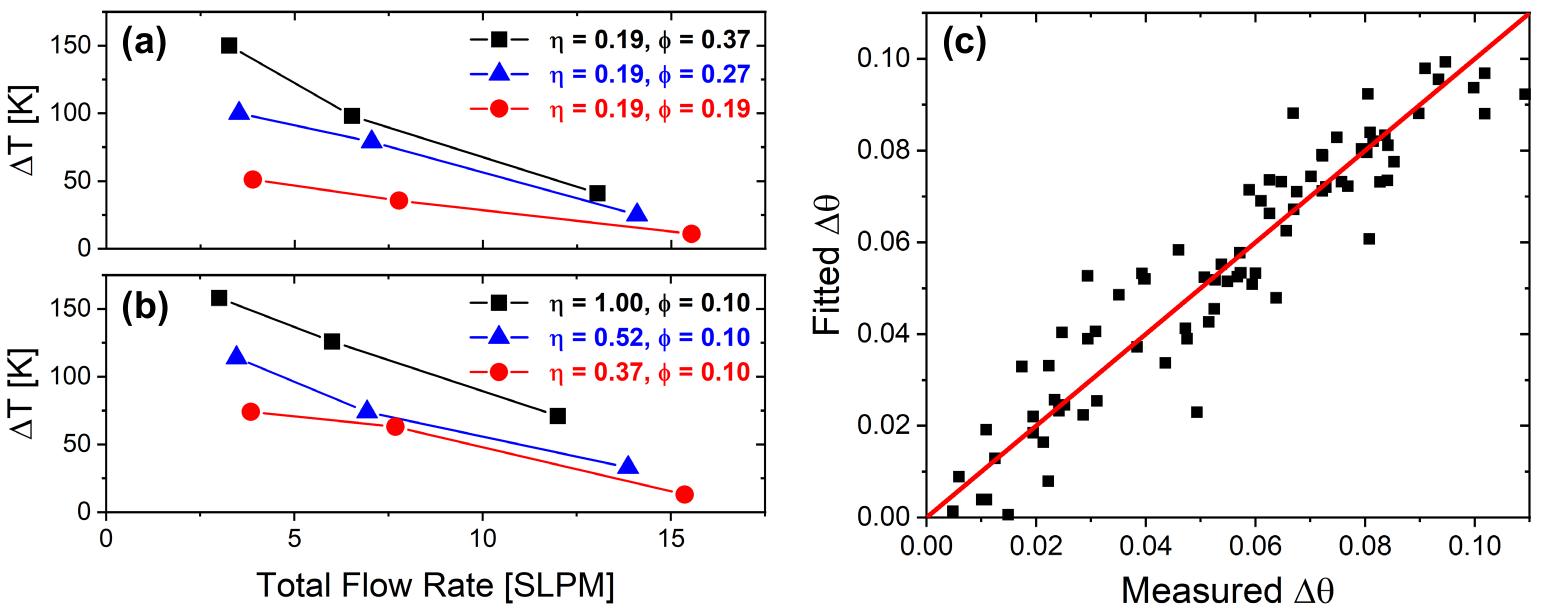}
\caption{Reduction in the burned gas temperature under various flame conditions. (a) At a fixed equivalence ratio $\phi$, $\Delta T = T_{b,ad} - T_b$ increases with the dilution factor $\eta$ and decreases with the total flow rate. (b) At a fixed dilution factor $\eta$, $\Delta T$ increases with $\phi$ and decreases with the total flow rate. (c) Comparison of the empirical fit for $\Delta \theta = 1 - T_{b}/T_{b,ad}$ with the measurement data.}
\label{fig4}
\end{figure}

The reduction in the burned gas temperature due to heat loss, $\Delta T = T_{b,ad}-T_b$, varies significantly with mixture composition and flow velocity, as illustrated in Fig. \ref{fig4}. An empirical fit for the non-dimensional temperature change, $\Delta\theta = 1-T_b/T_{b,ad}$, as a function of the equivalence ratio $\phi$, the dilution factor $\eta = \dot{m}_{\rm H_2} / (\dot{m}_{\rm H_2} + \dot{m}_{\rm N_2})$, and the non-dimensional flow speed $\xi = V_u/S_{L,ad}$, is shown below:

\begin{equation}
\begin{aligned}
\Delta\theta = 0.081\eta^{-0.12}\phi^{-0.15}exp(-0.40\xi).
\end{aligned}
\end{equation}

\subsection{Numerical Simulation of the Cellular Flame Structure}
The current numerical simulations were conducted using the EBIdnsFOAM code, developed by Bockhorn, Zirwes and co-workers \cite{bockhorn2012implementation, zirwes2023assessment}, on the OpenFOAM platform \cite{weller1998tensorial}. A multi-component diffusion model including detailed Soret diffusion was employed \cite{schlup2018validation}, and the chemical reactions were modeled using finite-rate chemistry based on the 9-species H$_2$-O$_2$ mechanism developed by Li et al. \cite{li2004updated}.

\begin{figure}[ht!]
\centering
\includegraphics[width=0.5\columnwidth]{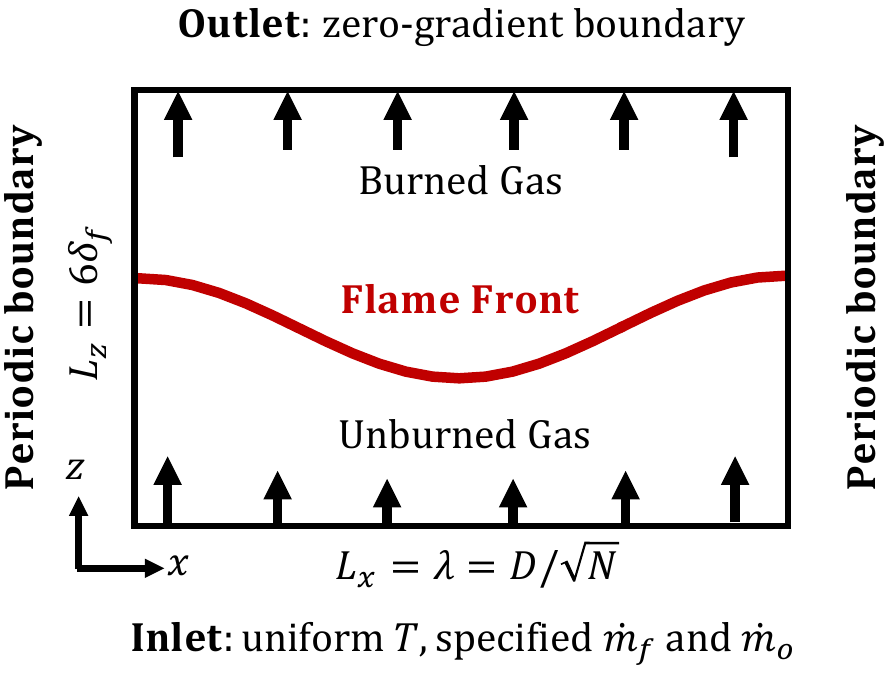}
\caption{A schematic of the current computational setup and boundary conditions. At the inlet, the mass flow rates of individual species (i.e., $\dot{m_k}= \int_0^{L_x} (\rho U Y_k -\rho D_k \partial Y_k /\partial z - D_k^T \partial \ln T/\partial z) dx$, where the three terms represent convection, mass diffusion and Soret diffusion, respectively) were specified according to the nominal experimental conditions.}
\label{figdomain}
\end{figure}

Fig. \ref{figdomain} illustrates the computation domain and boundary conditions of the current numerical simulations. The horizontal length of the computational domain was set as the average cell size ($\lambda = D/\sqrt{N}$, with $D$ = 18 mm being the burner diameter and $N$ being the total number of cells) determined from the measurement data. The vertical length of the domain was set to six times the flame thickness, ${\delta_f} = (T_{max} - T_{min})/|\nabla T|_{max}$, which was calculated from a one-dimensional burner-stabilized flame at the nominal experimental conditions. To accurately resolve the flame structure, the computation domain was uniformly discretized with a grid spacing of $\Delta x = \Delta z = \delta_f/64$.

An inlet boundary condition was assigned to the bottom boundary of the computation domain, with the mass flow rates of fuel and oxidizer specified according to the experimental conditions. At the top boundary, zero-gradient conditions were applied to the species mass fractions, velocity, and gas temperature. Periodic boundary conditions were imposed on both the left and right sides of the domain. 

Initial conditions were assigned based on the mass flow rates of H$_2$, O$_2$ and N$_2$, and the actual surface temperature measured experimentally. Under conditions of relatively high equivalence ratios, the streamwise distributions of flow velocity, gas temperature, and species mass fractions were initialized using the one-dimensional burner-stabilized flame solution obtained from Cantera simulations, whereas the spanwise distributions were set as uniform. To trigger the growth of cellular instability, a small perturbation was imposed on the initial conditions by slightly curving the flame front; any unperturbed scalar field $A_0(x,z)$ was shifted along the flow direction by a sinusoidal waveform, yielding $A(x,z) = A_0(x,z + a sin(2\pi x/\lambda))$. In the current study, the amplitude of the initial perturbation was set to $a = \delta_f/64$.

For very lean mixtures, the nominal laminar flame speed under adiabatic conditions ($S_{L,ad}$, for uniform distributions) was lower than the total flow speed of the unburned gas ($U$), and direct initialization using relatively uniform spatial distributions would lead to numerical blow-off. To circumvent this problem, an iterative initialization protocol was adopted. First, a much higher equivalence ratio, for which $S_{L,ad}>U$, was selected as the base case, and for this case a cellular flame simulation was initialized using the aforementioned method. The simulation was then advanced in time until steady-state distributions of flow velocity, gas temperature, and species mass fractions were achieved. Next, these steady-state distributions were used to initialize another simulation at a low $\phi$ but the same $\eta$ by linearly scaling the species mass fractions and flow velocity based on the modified oxygen flow rate. In small steps, $\phi$ was progressively reduced to the target value, with each new simulation initialized from the steady-state distributions of the previous one. This protocol ensured a robust and efficient transition to the target flame conditions, typically within two or three iterations. 

\section{Results and Discussion}
\subsection{Three-Dimensional Temperature and OH Concentration Fields}

\begin{figure}[ht!]
\centering
\includegraphics[width=\columnwidth,trim = 0 0 -2 0,clip]{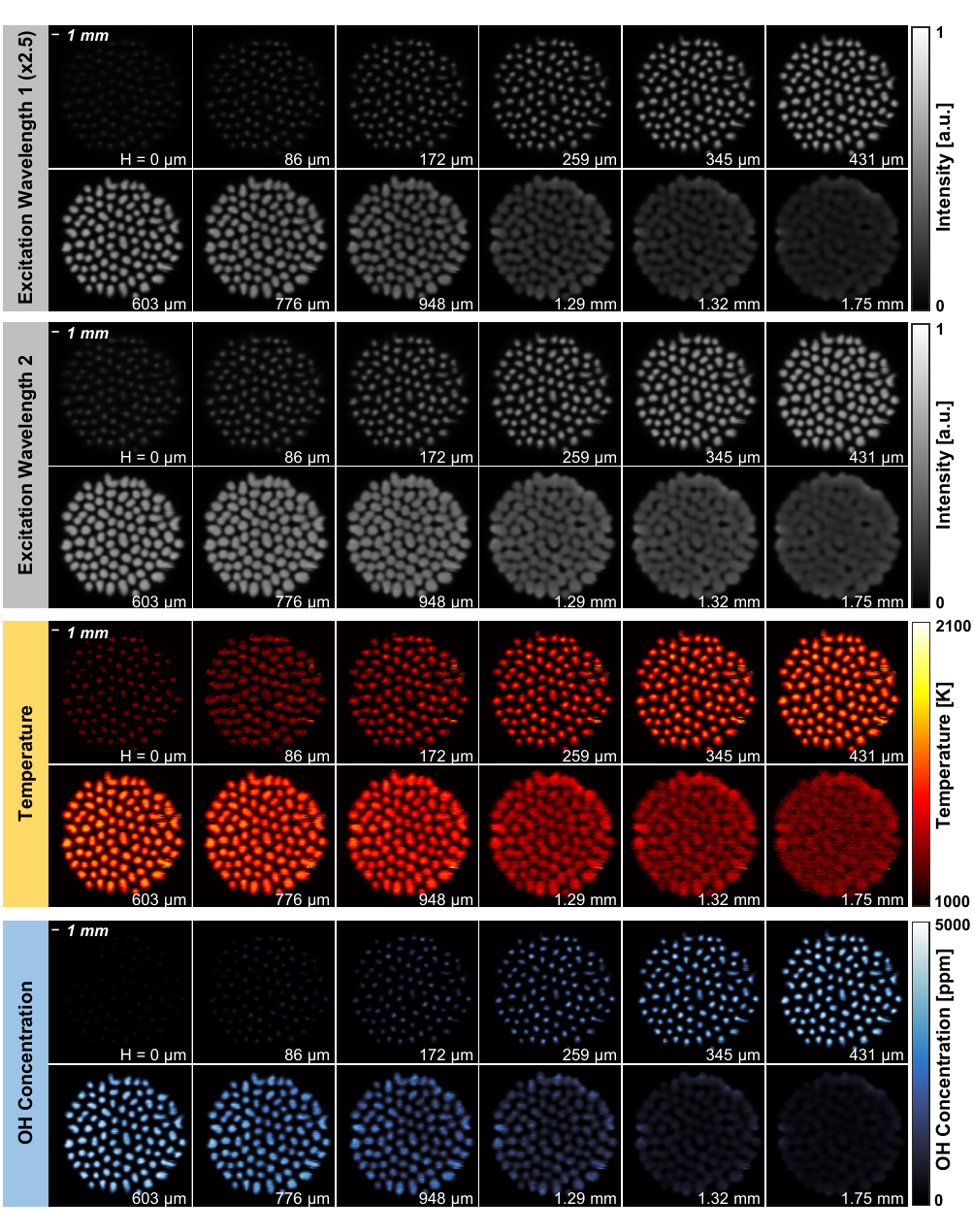}
\caption{Representative results for a cellular hydrogen flame under the conditions of $\dot{m}_{\rm H_{2}}$ = 1.00 SLPM, $\dot{m}_{\rm N_2}$ = 1.68 SLPM, $\dot{m}_{\rm O_2}$ = 5.00 SLPM. From top to bottom, the figure shows the horizontal distributions of OH-PLIF intensity at 282.952 nm and 282.997 nm excitation, followed by the inferred temperature and OH concentration, respectively. Results are shown for different heights above the burner (HAB).}
\label{fig6}
\end{figure}

Fig. \ref{fig6} shows results from a representative measurement for a lean hydrogen flame at $\phi$ = 0.10 and $\eta$ = 0.37. Note that these results correspond to horizontal slices of the 3D cellular flame, reconstructed from vertical slices obtained directly from the linear-translation measurements. The high contrast of these images indicates the quality of the current data. Based on these data, the spatial distributions of OH concentration and gas temperature were quantitatively determined.

In this example, individual cells or pockets of hot combustion products were clearly observed near the burner surface. The cells expanded downstream of the flame front and merged at a HAB greater than 2.0 mm. The burner surface temperature was kept at 302 K, creating a heat sink that helped stabilize the cellular flame. The complex flame structures resulted from competition among (a) local flame acceleration/deceleration induced by stretch and differential diffusion of mass and heat, (b) local flow acceleration/deceleration, and (c) flame deceleration induced by heat loss to the burner surface (which scales with the vertical gradient of the local gas temperature at the surface). Compared with freely propagating flames, where the cellular structures continuously vary with time, the structures of the current burner-stabilized flames remained steady throughout the measurement.

\clearpage

Under the conditions of Fig. \ref{fig6}, the flame contained 86 cells, and the dominant horizontal length scale of the cell structure was found to be approximately 2 mm. The average flame height above the burner, defined as the position of maximum total OH-PLIF intensity (under 282.997 nm excitation) across the horizontal plane, was approximately 0.9 mm.

\subsection{Dominant Wavenumbers of Cellular Flames}
\begin{figure}[ht!]
\centering
\includegraphics[width=0.5\columnwidth]{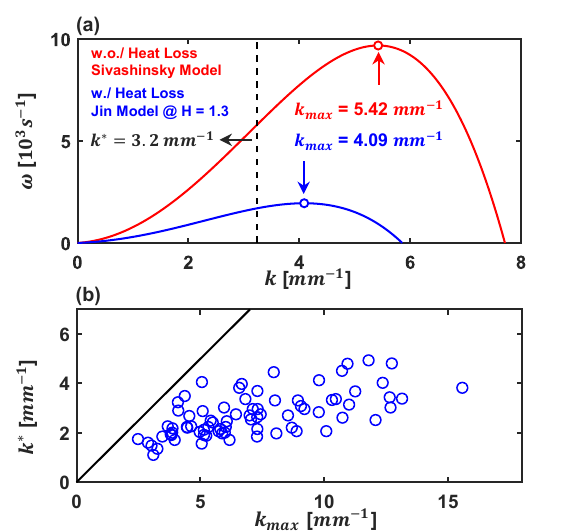}
\caption{Comparison between the dominant wavenumber at steady state ($k^*$) and the most unstable wavenumber predicted by dispersion relations ($k_{max}$). (a) Dispersion relations for the case shown in Fig. \ref{fig6}. Red line: the adiabatic Sivashinsky model \cite{Sivashinsky1977nonlinear}; blue line: the Jin et al. model \cite{jin2015experimental} with a heat loss factor of $H = 1.3$; black dashed line: the dominant wavenumber at steady state. (b) Summary of $k^*$ versus non-adiabatic $k_{max}$ for all cellular flames measured in the current study. The black line corresponds to $k^* = k_{max}$.}
\label{fig7}
\end{figure}

The dominant wavenumber of a cellular flame can be estimated based on the average cell size, i.e., $k^* = 2\pi/\lambda = 2\pi\sqrt{N}/D$. For example, the case shown in Fig. \ref{fig6} corresponds to a dominant wavenumber of $k^*$ = 3.24 mm$^{-1}$. This value is further compared with the most unstable wavenumber predicted by the dispersion relations with and without heat loss, as shown in Fig. \ref{fig7}. For this particular case, the relevant parameters involved in the dispersion relation calculation are: $\sigma$ = 4.02, $Ze$ = 19.0, $Le$ = 0.412, $S_L$ = 0.117 m/s, $D_T$ = $3.31 \times 10^{-5}$ m$^2$/s, and $H = 1.3$. 

As Fig. \ref{fig7}(a) clearly indicates, introducing heat loss to the flame can reduce the growth rate of cellular instability and narrow the range of unstable wavenumbers, which agrees with experimental observations that heat loss enhances flame stability. The most unstable wavenumber corresponding to the maximum growth rate ($k_{max}$) also decreases with heat loss. A summary of all cellular flame measurements in the current study, shown in Fig. \ref{fig7}(b), reveals that the dominant wavenumber of a steady-state cellular flame is consistently lower than the most unstable wavenumber predicted by the linearized dispersion relation. This observation indicates that nonlinear interactions between finite-amplitude perturbations at different length scales favor the growth of low-frequency components at long times. A similar trend has been observed in recent numerical studies on freely propagating lean hydrogen flames; for example, in the study by Zirwes and co-workers \cite{zirwes2024role}, flame cells much larger than the initial perturbation are formed, and the spatial mode of the fastest flame surface area growth rate shifts to longer wavelengths as time evolves. On the experimental side, to the authors' knowledge, the present study provides the first three-dimensional measurements of the ultimate statistics of burner-stabilized cellular flame morphology at steady state. Further details on the spatial distributions of key scalars within the cell structure are discussed in the next section.

\subsection{Scalar Distributions in a Representative Flame Cell}

\begin{figure}[ht!]
\centering
\includegraphics[width=1\columnwidth]{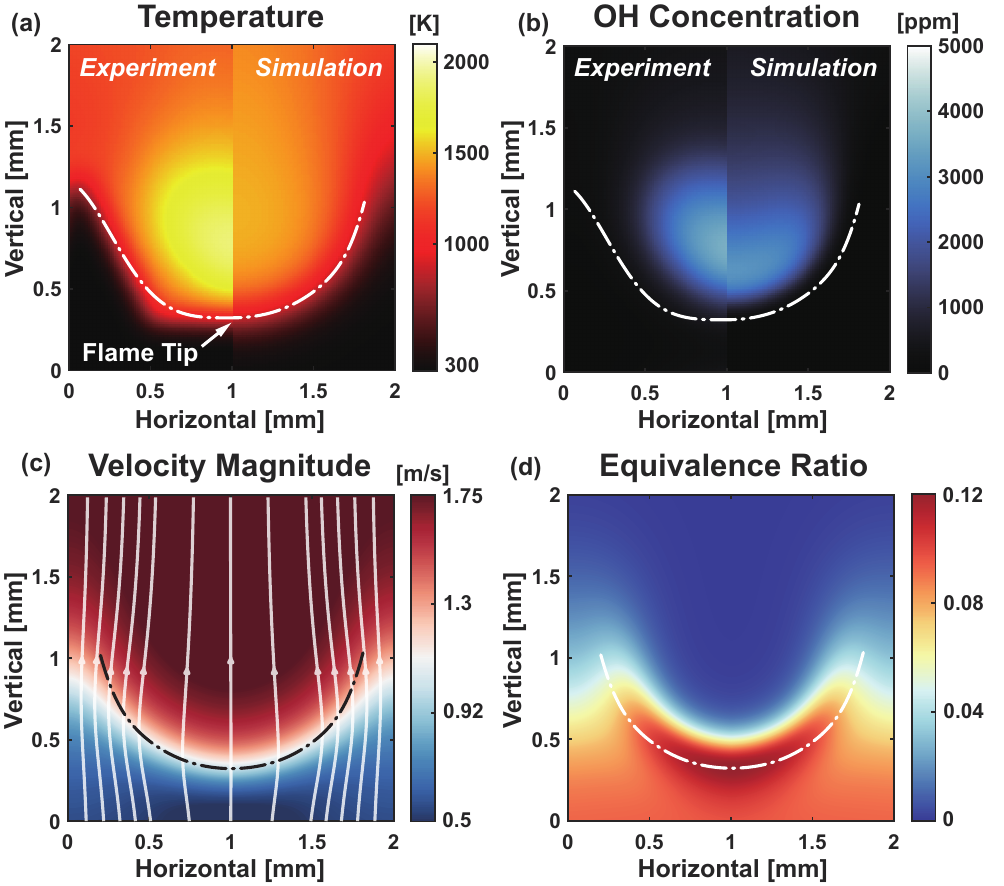}
\caption{Distributions of key scalar fields in a representative flame cell under the conditions of $\dot{m}_{\rm H_2}$  = 1.00 SLPM, $\eta$ =0.37, $\phi$ = 0.10. (a)-(b) Comparison between the measured and simulated distributions of temperature (left) and OH mole fraction (right) within the central cross-section of the flame cell. Qualitative agreement is observed between the current measurements and simulations. Dash-dot lines indicate the flame front locations determined from the magnitude of the temperature gradient. A region of super-adiabatic temperature is observed downstream of the flame front. (c) Flow velocity distribution in the flame cell, shown together with several representative streamlines. Flow expansion near the curved flame tip creates a relatively low-speed region that helps anchor the flame. (d) Spatial variations of the simulated equivalence ratio, evaluated from the local fluxes (including contributions from convection and diffusion) of hydrogen and oxygen. Local enrichment of hydrogen is observed upstream of the flame tip. }
\label{fig8}
\end{figure}

Fig. \ref{fig8} shows the spatial distributions of several key scalar fields (gas temperature, OH mole fraction, velocity magnitude and equivalence ratio) in a representative flame cell at nominal flow conditions of $\dot{m}_{\rm H_2}$  = 1.00 SLPM, $\eta$ = 0.37, $\phi$ = 0.10. For the temperature and OH concentration distributions, both the measurement and simulation results are presented, and they agree qualitatively (with some minor differences probably caused by three-dimensional effects), demonstrating the high fidelity of the current simulation. As for the velocity and equivalence ratio distributions, only the simulation results are available. Under these conditions, the nominal adiabatic flame speed ($S_{L,ad}$ = 0.12 m/s) is significantly less than the average flow velocity of the unburned gas mixture ($V_u$ = 0.50 m/s), which would have led to flame blow-off if cellular instabilities were not present. It is evident that the cellular structure plays a critical role in stabilizing the flame.

Further analysis of these scalar fields reveals three key effects contributing to the stabilization of the cellular flame. The first is flame acceleration induced by curvature. For lean hydrogen mixtures, the corresponding Markstein length is negative \cite{aung1997flame}, which means that a positive flame stretch increases the laminar flame velocity. From a Lewis-number perspective, at the flame tip where the local stretch is positive, the rapid diffusion of hydrogen (as compared to thermal diffusion) creates a focusing flux of energy that exceeds heat conduction to the preheat zone, thereby strengthening the flame. This focusing flux of energy also generates a small region downstream of the flame front where the local gas temperature exceeds the adiabatic temperature defined by the nominal mixture composition, as evident in Fig. \ref{fig8}(a). This super-adiabatic behavior has been reported in various numerical studies (for example, see \cite{berger2019characteristic, zirwes2024role}); however, direct experimental observation has been scarce prior to the present study. In this region, the measured OH concentration is also seen to exceed its nominal equivalence value, as illustrated in Fig. \ref{fig8}(b).

The second effect, also related to the curved flame front, is the modulation of local velocity distribution by flow expansion and compression. As shown in Fig. \ref{fig8}(c), the convex tip of the flame near the burner surface expels the flow sideways, distorting the streamlines and creating a region upstream of the flame front where the local unburned gas velocity is much lower than the average value. The flame is stabilized at locations where the local laminar flame speed matches the projected flow velocity normal to the flame front. In the current study, the exact location of the flame front, as indicated by the dash-dot lines in Fig. \ref{fig8}, is defined by a curve passing the point of maximum temperature gradient ($|\nabla T|_{max}$) such that, along its normal direction, the gradient of the magnitude of temperature gradient is zero: $\partial|\nabla T|/\partial n = 0$.

Note that the two previously described effects generally exist in any flame with a Lewis number less than unity. The third effect, however, is specific to very light fuel species such as hydrogen. This effect arises from thermodiffusion or the Soret effect, which is caused by the temperature dependence of molecular diffusivity that drives hydrogen toward hotter regions such as the flame front. This preferential diffusion of hydrogen toward higher temperature, illustrated in Fig. 8(d), creates a local enrichment of the unburned gas mixture by increasing its effective equivalence ratio. Near the flammability limit, this enrichment leads to a pronounced increase in the laminar flame speed, which further aids flame anchoring against the incoming flow.

\clearpage

\subsection{Regime Diagrams of Lean Hydrogen Flames}
\begin{figure}[ht!]
\centering
\includegraphics[width=1.0\columnwidth,trim = 0 0 -2 0,clip]{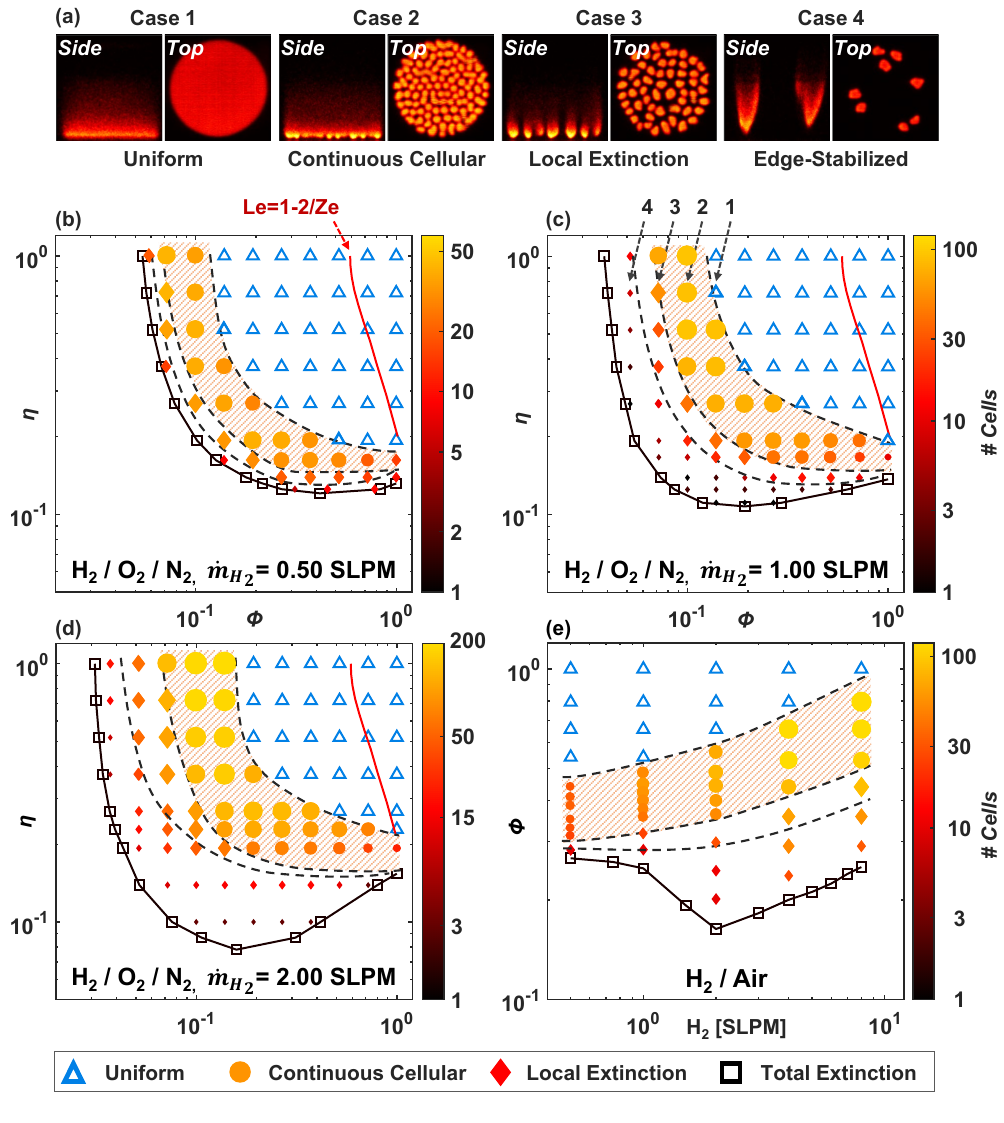}
\caption{Regime diagrams for the morphology of lean hydrogen flames measured in the current study. (a) Representative examples of flame morphology in four different regimes. (b)-(d) Flame measurements at H$_2$ flow rates of 0.50, 1.00, and 2.00 SLPM, respectively, with $\phi$ and $\eta$ varied independently. (e) Measurements of H$_2$-air flames at various H$_2$ flow rates and $\phi$. Open triangles: stable flames; filled circles: continuous flames subject to cellular instability (with the number of cells represented by the size and color scale of the symbols); diamonds: cellular flames with local extinction; open squares: boundary cases of total flame extinction/blow-off. The theoretical limit for the onset of cellular instability in adiabatic flames, determined by the critical Lewis number criterion, $Le = 1 - 2/Ze$, is indicated by the red curve. }
\label{fig9}
\end{figure}

In the current study, a total of 279 measurements of lean hydrogen flames were conducted at H$_2$ mass flow rates of 0.50, 1.00, and 2.00 SLPM. The equivalence ratio $\phi$ and the dilution factor $\eta$ were varied independently -- for any fixed $\eta$, $\phi$ was scanned from unity to the extinction limit with equal steps in log space. The results are summarized in Fig. \ref{fig9}.

The flame appears stable and relatively uniform when $\phi$ and $\eta$ are both close to unity (Case 1). Cellular patterns emerge as either parameter decreases sufficiently from unity (Case 2). In the presence of a cold burner surface, the observed onset of instability is significantly delayed compared with the classic theoretical limit for adiabatic, free-propagating flames, i.e., $Le = 1 - 2/Ze$. The wrinkled flame front remains continuous when the cellular instability first appears, but as $\eta$ or $\phi$ decreases further, local extinction occurs along the flame front, segmenting the flame into isolated cells that cease to merge downstream (Case 3). Beyond this point, the number of cells drops rapidly. Further decreasing $\phi$ or $\eta$ pushes the individual flame cells to the edge of the burner, where the local flow speed is relatively slow and favors flame anchoring (Case 4). At sufficiently high dilution levels or sufficiently low equivalence ratios, the entire flame blows off completely, as denoted by the open squares in Fig. \ref{fig9}. 

The transition boundaries between these regimes are represented by dashed lines in Fig. \ref{fig9}. As the H$_2$ flow rate increases, the boundary of the stable flame regime shrinks monotonically. This is likely due to the reduced volumetric heat loss rate at higher flow rates, which lessens the stabilization effect of the burner surface. By contrast, the boundary between the continuous flame and local extinction remains nearly unchanged as the H$_2$ flow rate varies. Note that these two boundaries reflect intrinsic properties of lean hydrogen flames stabilized on a constant-temperature surface and are expected to be independent of the specific burner geometry. In contrast, the boundaries of edge-stabilized flame and total extinction depend on the burner diameter $D$. An asymptotic analysis for $D$ approaching infinity -- where edge effects vanish -- is reserved for future studies. In the present study ($D$ = 18 mm), these two boundaries expand monotonically with increasing H$_2$ flow rate over the range 0.50 - 2.00 SLPM, likely due to the reduction in volumetric heat loss rate. However, they are expected to contract at sufficiently high H$_2$ flow rates as the flame tends to blow off more easily. Further discussion on the quantitative determination of regime boundaries is provided in the Supplementary Material.

\clearpage

A total of 60 additional measurements were also conducted for hydrogen-air flames at various equivalence ratios and H$_2$ flow rates. The results are shown in Fig. \ref{fig9}(d). It is evident that the total extinction limit varies nonmonotonically with respect to the H$_2$ flow rate, reflecting a complex competition between the reduced volumetric heat loss rate (which tends to increase the laminar flame speed and stabilize the flame) and the increased flow velocity (which promotes blow-off). For experiments on the current burner, the blow-off equivalence ratio reaches a minimum value of $\phi_{b,min} = 0.16$ at an H$_2$ flow rate of 2.00 SLPM. In contrast, the critical equivalence ratios for both the onset of cellular instability and the occurrence of local extinction increase monotonically with the H$_2$ flow rate. Summary tables for all hydrogen cellular flame measurements conducted in the current study are provided in the Supplementary Material. The original experimental data are also available upon request via email.

\section{Conclusions \label{sec: conclusions}}
A systematic characterization of burner-stabilized lean hydrogen flames was performed across a wide range of equivalence ratios, dilution factors, and flow rates, with spatially resolved measurements of the three-dimensional temperature and OH concentration fields obtained using combined techniques of multi-wavelength OH-PLIF and laser sheet scanning. Based on experimental data from over 200 flame cases, regime diagrams for four flame modes -- stable flames, continuous cellular flames, cellular flames with local extinction, and edge-stabilized isolated flame cells -- were accurately determined.

Comparison with linear stability analysis revealed that the dominant wavenumbers of steady-state cellular flames were significantly lower than the most unstable wavenumbers predicted by the linearized dispersion relation, indicating that nonlinear interactions between finite-amplitude perturbations of different length scales favored the growth of low-frequency components at long times. The cell structure played an important role in anchoring the flames near the burner surface, especially at nominal equivalence ratios near the lean flammability limit, where the flames would have blown off without them. The mechanism of cellular flame stabilization, as revealed by complementary numerical simulations using a detailed reaction model, was three-fold: (1) positive curvature at the tip of the flame front accelerated the local flame speed due to the negative Masktein length of a lean hydrogen mixture; (2) the unburned gas velocity distribution was modulated by local expansion/compression ahead of the curved flame front near the burner surface; and (3) local stratification of equivalence ratio occurred as a result of Soret diffusion. Together, these effects created regions of lower speed and richer mixture that sustained flame anchoring. The findings and new data from the present study are expected to advance fundamental understanding of flame dynamics in lean hydrogen mixtures and to provide guidance for the design of practical hydrogen combustors.

\section*{Acknowledgments}
This work was supported by the National Natural Science Foundation of China under Grants No. 92152108 and No. 12472278, and the Space Application System of China Manned Space Program under the projects "Nonlinear Dynamics of Flame Instability under Microgravity Conditions" and "Ignition Mechanism of Premixed Near-Limit Flames under Microgravity Conditions". Numerical simulations were supported by the High-Performance Computing Platform of Peking University.

\bibliographystyle{elsarticle-num-names}
\bibliography{References}

\end{document}